\documentstyle[preprint,eqsecnum,aps]{revtex} 
 
\begin{document} 
\draft 
\preprint{} 
\title{A topological characterization of delocalization in a spin-orbit 
coupling  system} 
\author{D.N. Sheng and  Z. Y. Weng} 
\address{Texas Center for Superconductivity and Department of Physics\\ 
University of Houston, Houston, TX 77204-5506 }  
\maketitle 
\begin{abstract} 
We show that wavefunctions in a two-dimensional (2D) electron system  
with  spin-orbit coupling can be characterized by a topological quantity--the 
Chern integer due to the existence of the intrinsic Kramers degeneracy.  
The localization-delocalization transition in such a system is studied in terms
of such a Chern number description, which reproduces the known metal-insulator 
transition point.  The present work suggests
a unified picture for various  known 2D delocalization phenomena based on 
the same topological characterization. 
\end{abstract}

\vspace{0.8in} 
PACS numbers:  71.30.+h, 72.15.Rn, 71.70.Ej

\newpage  
Metal-insulator phase transition problem in two dimensions (2D) has attracted a
lot of attention in recent years. With   time-reversal and spin rotational 
symmetries, all states in a 2D non-interacting
electron system should be localized due to disorders, according to the Anderson
localization.$^{1,2}$ In contrast to such so-called orthogonal systems, 
extended states exist in  quantum Hall effect (QHE) systems$^{3,4}$ to carry 
current going through whole samples. These systems in which the time-reversal 
symmetry  is broken by  external  
magnetic fields  are known as the unitary class. There are still on going 
debates$^{5-7}$ about whether 
localization-delocalization transition could happen in a  random magnetic field
case.  For a time-reversal invariant spin-orbit coupling system,  analytic and 
numerical  studies$^{8-10}$ have been given consistent  
conclusions about the existence of a  metal-insulator transition in this 
so-called symplectic class. 
 
These 2D quantum systems have been  classified as three different universality
classes,$^{11}$ namely, orthogonal, unitary, and symplectic,  in terms of 
symmetries of  Hamiltonians. It is generally believed that symmetries of these 
systems may play important roles in determining different
localization-delocalization properties. But it is an interesting open
question whether  2D localization-delocalization transitions in different
symmetry classes are intrinsically
different,  or, all of them (both unitary and symplectic classes) are
due to the same universal reason?
 
In the presence  of magnetic fields,  Thouless and co-workers$^{12}$ and 
others $^{13,14}$ 
have shown that the delocalization property of a wavefunction is well 
characterized by its associated Chern 
number (which is equal to  the quantized-Hall conductance in the unit of 
$e^2/h$). As a topological 
invariant integer, Chern number represents a non-trivial topology of 
wavefunction. It has been shown$^{14}$  that nodes of an eigenstate 
wavefunction with nonzero Chern number can move freely to cover the whole real 
space when one continuously  changes the boundary 
condition.  Thus a nonzero Chern number describes  the extensiveness of 
a wavefunction. In contrast, a state with zero Chern number should always be  
localized in 2D due to  Anderson localization.$^{12}$  Numerical 
calculations$^{15}$ based on Chern numbers 
have excellently verified localization-delocalization phase diagram in the QHE 
system. In  random-magnetic-field systems,  even though the total Hall 
conductance 
becomes zero on average, nonzero Chern integer (or quantized Hall conductance) 
for single-electron states can still exist, and recently a metal-insulator 
transition  has been revealed$^7$ numerically in terms  of the classification 
of Chern integers. 
 
It is natural for one to ask if the delocalization property in spin-orbit  
coupling systems can be similarly understood in terms of such a  topological 
description. 
Without uniform or random magnetic field, one may expect a zero Chern number 
and thus a trivial topology for wavefunctions in their parameter space.   
However,   this is not necessarily correct. According to Berry,$^{16}$  when one  
adiabatically changes a wavefunction in its parameter space, the Berry phase of 
the wavefunction can exhibit a  monopole-like topological structure if there is 
a two-fold degeneracy. In the spin-orbit coupling system,  
there is a well-known two-fold Kramers degeneracy of eigenstates (even in the  
presence of random scattering) due to the time-reversal invariance. In this  
paper, we analytically show that such a degeneracy indeed leads to a 
non-trivial topological classification of wavefunctions in a spin-orbit 
coupling system which is essentially the same as in the unitary class cases,  
and thus one can relate delocalization  
property in this system to its topological characteristics. Numerically we 
demonstrate that a topological 
transition happens at the point where the density of nonzero Chern numbers 
become vanishing when the 
system is extrapolated to  
infinity, and such a transition indeed coincides with the known critical point 
of metal-insulator  
transition obtained based on conventional numerical methods.$^{10}$  
Therefore, a unified picture underlying the localization-delocalization  
transition in 2D systems (both unitary and symplectic classes) is established  
where all the delocalizations can be essentially characterized by the same
topological quantity-- the Chern integer. 
 
We start with a tight-binding lattice model of noninteracting electrons  
with spin-orbit coupling. The Hamiltonian is defined as follows:$^{10}$ 
$$H=-t\sum_{\tau=\hat {x},\hat {y}} \sum _{i} \hat {c}_i^+\hat {c}_{i+\tau}    
    -iV \sum_{i }\hat {c}_i^+\sigma _y\hat {c}_{i+\hat {x}}  
    +iV \sum_{i }\hat {c}_i^+\sigma _x \hat {c}_{i+\hat {y}} 
    +\frac 1 2 \sum _i w_i \hat {c}_i^+\hat {c}_i    + H.c. 
\eqno (1) $$ 
Here $\hat {c}_i^+=(c^+_{i\uparrow},c^+_{i\downarrow})$ are  fermionic 
creation operators,  with $t$ as 
the nearest-neighbor hopping integral and  V as hopping-spin-flipping  (i.e., 
spin-orbit coupling)  term.   
The strength of  spin-orbit coupling is  represented by a dimensionless  
parameter $S=V/\sqrt {t^2+V^2}$. 
And $w_i$ is a random potential with strength $|w_i|\leq W/2$ ($W$ is 
adjustable). For simplicity, 
we assume no correlations  among different sites for $w_i$ (i.e., white noise 
limit). Such a system will be studied  under a generalized boundary condition: 
$\hat {\Psi }(i+L_j)=e^{i \theta _j}\hat {\Psi} (i) $ with square lattice 
width $L_1= L_2=L$ and a total number of lattice sites  
${N}=L \times L$ (here $\hat {\Psi }$ includes two spin components and   
j=1,2 represent x and y directions, respectively). 
 
Eigenstates of $H$  may be regarded as functions of boundary-phases $\theta_1$  
and $\theta_2$, and  topological property of each eigenstate (labeled by $m$)
can be characterized in terms of relations with $\theta_1$ 
and $\theta_2$ through the Chern integer defined as follows:$^{12-14}$ 
$$ C^{(m)}=\frac {i} {2\pi} \int\\ \int d \theta _1 
d \theta_2 \sum_{i}    F(\theta _1, \theta _2; i),   \eqno (2a) $$ 
in which  
$$ F(\theta _1, \theta _2, i)= 
 \left\langle \frac {\partial \hat {\Psi} _m^+ (\theta_1,\theta_2;i)} 
{\partial \theta _1}\left|\frac {\partial \hat {\Psi} _m 
(\theta_1,\theta_2;i)}   
{\partial \theta _2} \right\rangle\right.  
  -\left\langle \frac {\partial \hat {\Psi} ^+ _m (\theta_1,\theta_2;i)}  
 {\partial \theta _2} \left|  
\frac {\partial \hat {\Psi} _m 
(\theta_1,\theta_2;i)} {\partial \theta _1 } \right\rangle\right. , 
  \eqno (2b)$$ 
and the area-integral in (2a) covers a $2\pi \times 2\pi $ unit cell in the  
$\theta$-space. The Chern number $C^{(m)}$ in (2a) can be shown$^{12,13}$ to be 
an integer.  Naively one would expect a zero Chern number $C^{(m)}$  
for a  system with time-reversal symmetry, since $C^{(m)}$  is  
proportional to a Hall conductance as mentioned earlier. In fact, one would see 
cancellation in (2) due to the symmetry in wavefunction $\hat{\Psi}$: 
$\left( \begin{array}{c}  \Psi _{\uparrow}  
(\theta_1,\theta_2;i) 
\\                                       
\Psi _{\downarrow} (\theta_1,\theta_2;i) 
\end{array}\right )     
=\left ( \begin{array}{r}  \Psi _{\downarrow}^* (-\theta_1,-\theta_2;i) 
\\                                       
-\Psi _{\uparrow}^* (-\theta_1,-\theta_2;i)  
\end{array}  \right )   $, which  reflects the time-reversal invariance of the  
original  Hamiltonian. However, in the spin-orbit coupling case,  
there is  always a two-fold degeneracy (Kramers degeneracy) in Hamiltonian (1) 
at boundary conditions with $\theta_1$, 
$\theta_2 = \pi  \times $ integer.$^{10}$  Both $\hat{\Psi}_m$ and its 
time-reversal counterpart state are degenerate and orthogonal here. 
Due to such a degeneracy, derivatives in (2b) would become uncertain at the
degeneracy points and the contribution from these points needs more careful
examination. 

In fact, the Chern integer $C^{(m)}$ is equivalent to a Berry phase (divided by 
$2\pi$) defined on a closed path along the boundary of a unit cell,$^{16}$ 
which may be converted to a surface
integration like in (2a) by Stokes theorem. Since each degeneracy point
in the parameter space will lead to a monopole-like singular contribution to 
the Berry phase,$^{16}$
a 2D $\theta$-plane would cut right through such a singularity point. To
avoid this difficulty, one has to introduce a third parameter $\epsilon$ to 
deform the surface integral
around the degeneracy points. For example, we may add an  
infinitesimal term $H_p=-\epsilon \sum_i \hat {c}_i^+\sigma _z 
\hat {c}_i$ to the Hamiltonian. Then
the surface integration in (2a) is understood as the one in a
three-dimensional parameter space ($\theta_1$, $\theta_2$, $\epsilon$)
(here $\epsilon\ll \mbox{max} (t, V, w_i)/N$ so that
it has no effect when one is sufficiently far away from the degeneracy points).
Due to the cancellation in the regions far away from the degeneracy points,
Eq.(2a) is then reduced  to 
$$ C^{(m)}=\frac {i} {2 \pi  } \sum_{k=1}^4 \int\\  \int _{S_k} 
d \theta _1 d \theta _2 \sum 
_{i} F(\theta _1, \theta _2; i)     \eqno (3)   $$ 
in which  $S_k$ ($k=1-4$) represent four small surface-integrals covering the  
degeneracy points (0,0), ($\pi$, 0), (0, $\pi$), and ($\pi$, $\pi$) within a
$2\pi\times 2\pi$ unit cell. The minimum size of these areas is decided by 
the strength $\epsilon$ and the final result should not depend on
$\epsilon$ at $\epsilon\rightarrow 0^+$. One may find that this regularization 
procedure resembles the one in a magnetic phase transition where an
infinitesimal external symmetry-breaking field is usually introduced to lift  
the ground-state degeneracy for a spontaneous magnetization. Here $H_p$ plays a
role to lift the Kramers degeneracy such that the hidden topological
characterization becomes explicitly shown (see below).
 
Denote each degeneracy point as ($\theta^0_{1k}$, $\theta^0_{2k}$) ($k=1-4$)
and suppose that eigenstates 
are already known at these points for a given $\epsilon$. 
Then one can determine eigenstates at an arbitrary boundary phase ($\theta_1$, 
$\theta_2$), which  
is close enough to  ($\theta^0_{1k}$, $\theta^0_{2k}$), by a perturbative
way. The following procedure is similar to that in Ref. 17 in treating
the level-crossing problem.  The
change of the boundary condition may be incorporated  into the Hamiltonian as a
perturbation, and in terms of the complete set of basis obtained  at 
$(\theta_{1j}^0,\theta_{2j}^0)$,  such a change in Hamiltonian can be 
generally written as 
$$\Delta H(\theta_1,\theta_2)=   \left( \begin{array}{clcr}   
0 & \Delta _{n}^*    
\\                                       
 \Delta _{n} & 0 
\end{array} \right ),    
     \eqno (4)  $$     
where $2\times 2$ matrix elements are related to the (2n-1)-th and (2n)-th 
eigenstates (assuming an energy-increase order for all the eigenstates), which  
correspond to a pair of two-fold degenerate eigenstates of $H$. 
The off-diagonal terms may be expanded to a leading order of 
$(\theta_1-\theta_{1k}^0)$  and $(\theta_2-\theta_{2k}^0)$ (If the leading term
happens to be zero, higher order expansions in $\Delta H(\theta_1,
\theta_2)$ will be needed, and integer Chern numbers can be similarly 
obtained$^{17}$). The general form is 
$\Delta _n(\theta _1, \theta _2)=\alpha _k(\theta_1-\theta_{1k}^0)+\beta_k 
(\theta_2-\theta_{2k}^0)  $ 
with $\alpha_k=\frac {\partial }{\partial \theta _1} 
<\Psi_{2n-1}(\theta_{1k}^0, \theta_{2k}^0)|H+H_p|\Psi_{2n}(\theta_{1k}^0,  
\theta_{2k}^0)>$ 
and $\beta_k=\frac {\partial }{\partial \theta _2} 
<\Psi_{2n-1} (\theta_{1k}^0, \theta_{2k}^0)|H+H_p|\Psi_{2n}(\theta_{1k}^0,  
\theta_{2k}^0)> $. By diagonalizing the $2\times 2$ matrix of the
Hamiltonian for each n, we get the (2n-1)-th and 
(2n)-th eigenstates at the boundary phase  $(\theta _1, \theta _2)$  
sufficiently close to  $(\theta _{1k}^0, \theta _{2k}^0)$. Then according to  
(3), the Chern numbers for the (2n-1)-th and (2n)-th eigenstates are   
given by the following integers:$^{17}$ 
$$C^{(2n-1)}=\frac 1 2 \sum_{j=1}^4   {\mbox {sgn}} [\mbox{Im}
 (\alpha _j ^*\beta 
_j)] ;    \eqno(4a)$$        
$$C^{(2n)}=-C^{(2n-1)} ,   \eqno(4b)$$ 
whose values will not depend on $\epsilon=0^+$. (Here if $\mbox{Im}
(\alpha _j ^*\beta _j)=0$, we define $\mbox{sgn} [\mbox{Im} (\alpha _j ^*
\beta _j)]=0$ in (4a).)   
Therefore, eigenstates in the spin-orbit coupling system can be
indeed classified in terms of nontrivial Chern integers, and two degenerate 
eigenstates always have Chern numbers with opposite signs such that the total 
sum of them (and the sum of the Hall conductances) is still zero in consistence 
with the time-reversal symmetry.  Here each degeneracy 
point gives rise to a contribution similar to a magnetic monopole in unit 
strength 1/2 as originally discussed by Berry.$^{16}$    
Nonzero  Chern integers have been used to characterized extended  
states in both the QHE and random-magnetic-field systems.$^{14,15,7,18}$ By 
the same token, the delocalization problem in the spin-orbit coupling system  
should be similarly described here, after the Chern-number 
classification of wavefunctions is established. In the  
following, we shall use this topological characteristics to determine the  
localization-delocalization transition in the present system.    
 
We define a density of states $\rho _{ext} (\varepsilon, {N})$           
for the eigenstates with nonzero Chern number at energy $\varepsilon$.
$\rho_{ext}$ is related to the 
density of states for extended states.$^{15,7}$  
Exact diagonalization of $H$ is carried out at four boundary conditions where 
the two-fold degeneracy happens, and then Chern numbers 
are determined through the formula (4). The total density of  states  $\rho 
(\varepsilon , {N})$  and the extended one 
$\rho _{ext}(\varepsilon , {N})$ are calculated  as a function 
of  lattice size ${N}$ (${ 
N}$=36, 64, 144, 256 and 576), which is averaged over random potential  
configurations ($100-5000$ random configurations depending on sample sizes) and 
also averaged over a small energy  width $\Delta \varepsilon=0.4\sqrt{V^2+t^2}$
in the  neighborhood of energy $\varepsilon$  to reduce  statistics error. 
The ratio $\rho _{ext}(\varepsilon,{N})/ \rho ( 
\varepsilon, {N})$ at the band center $\varepsilon=0$ is presented in Fig. 1 
with a log-log plot. 
Different curves correspond  to different random strength $W$'s. 
The spin-orbit coupling strength $S$  has been chosen to be $S=0.5$ with $ 
\sqrt {V^2+t^2}=1$ (in order to compare with the results in Ref. 10).  
The ratio $\rho _{ext}(\varepsilon=0,{N})/ \rho ( 
\varepsilon=0, {N})$ monotonically decreases as parameter W increases, which 
reflects the fact that eigenstates become more and more localized at stronger 
disorders. For smaller $W$ ($W$=3, 4 and 5 as shown in the Fig. 1),  
the ratio $\rho _{ext}(\varepsilon=0,{N})/ \rho ( 
\varepsilon=0, {N})$ shows a flat curve with a slight increase at larger  
sample sizes,  suggesting that a finite value of 
$\rho _{ext}(\varepsilon=0,{N})/ \rho ( 
\varepsilon=0, {N})$  may be reached if it is extrapolated to  infinity 
lattice size. The trend becomes opposite at larger  strengths of random 
potential,  where the ratio   $\rho _{ext}(\varepsilon=0,{N})/ \rho ( 
\varepsilon=0, {N})$   becomes  monotonically decreasing  with larger sample 
sizes.  In fact , all the data 
at $W>5$ shown in Fig. 1 can be nicely fit into a straight line, which means  
$\rho _{ext}(\varepsilon=0,{N})/ \rho ( 
\varepsilon=0, {N}) \propto {N}^{-x}$ (with $x>0$ to be given below).  These 
finite-size scalings 
show that at a sufficiently large $W$, the density of states with nonzero 
Chern number  is always  
extrapolated to zero at large lattices. Similar to the cases in the QHE and 
random-magnetic-field systems, 
it means that all the eigenstates here have trivial topology and should be 
localized. 
 
The exponent $x$ as a function of the disorder strength $W$ is shown in Fig. 2.
It is interesting to see that all data fall on a straight line. Thus one can 
determine the critical strength  of disorder $W_c $  
at $x_c=0$ by extrapolation in Fig. 2.  Here $W_c$  decides a topological phase 
transition point. At $W>W_c$, $\rho_{ext}\sim N^{-x}\rightarrow 0 $ at
 $N\rightarrow \infty$, while at $W<W_c$, $\rho_{ext}\rightarrow $ a finite 
constant  at large lattice size. As have been known in the QHE and  
random-magnetic-field systems,  such a topological transition physically
 corresponds to a localization-delocalization transition.  In fact,
 $W_c$ determined from Fig. 2 
is approximately $5.62 \pm 0.08  $, which is quite close to the value $W_c=5.74$
obtained by conventional numerical methods$^{10}$ with much
larger lattice sizes $\sim 64\times 10^{5}$ (A slightly smaller 
$W_c$ in our result may be  due to the fact that  a finite energy average 
around $\varepsilon=0$ is made 
in calculation which could effectively lower $W_c$).  
 
The above analytic analysis and numerical calculation have shown 
that
delocalization occurs in the spin-orbit coupling case is basically the same as 
that in the unitary cases.  It suggests that all localization-delocalization 
transitions in 2D non-interacting systems may be universally described as a topological
phase transition in terms of the Chern integer. The conventional understanding 
of metal-insulator transition based on  symmetry classes of Hamiltonians may
not be  essential. For instance, by applying a small but finite magnetic field
one may change a spin-orbit coupling system from symplectic class to unitary
class by destroying the time-reversal symmetry. We have checked this case 
numerically and found$^{19}$ no sudden change in the distribution of Chern 
integers (thus the topological transition),  which actually behaves quite 
smoothly with small  magnetic fields. This fact has already been noticed before
in the behavior of localization length.$^{20}$ The universality has also been
shown in the study of the critical behavior of  metal-insulator transition,
where it has been found$^{21}$  that correlation  dimensions   
of   spectral measure  $\tilde {D}_2$ and  fractal eigenstate $ {D}_2$ as well 
as exponent $\eta$ describing the energy correlations of the critical 
eigenstates
in spin-orbit coupling system are very close to those obtained  
for the QHE systems.$^{22}$  Experimentally, both in   
random magnetic field$^{23}$ and   spin-orbit coupling$^{24,25}$ systems, 
magnetoconductances are negative in the delocalized region ( here 
``delocalization''  means the localization length is at least 
larger than the sample 
size), while positive in the localized region. 
This common feature can be understood as that in the delocalized region,  a
finite fraction of eigenstates carry nonzero Chern number and  magnetic 
fields suppress original phase coherence and reduce the number of 
states having nonzero Chern numbers.$^{19}$ Thus the magnetic field plays a 
role in reducing delocalization in  
the ``delocalized'' region and leads to negative magnetoconductance. In the
opposite case, where original wavefunctions are topology trivial, magnetic field
will be in favor of delocalization to give rise to a positive 
magnetoconductance.$^{26}$ This explains why  metal-insulator transition is
generally accompanied by a sign change of megnetoconductance experimentally in 
spite of  different symmetry classes.   
 
In conclusion, the localization-delocalization transition in a spin-orbit 
coupling system is studied from a topological point of view. It is shown that
wavefunctions in this time-reversal invariant system can  be characterized by 
Chern integers as a result of the intrinsic Kramers degeneracy. We have 
determined the critical  strength of disorder  $W_c$ for the metal-insulator 
transition point at energy $\varepsilon=0$ based on such a Chern-integer 
characterization, which is in excellent agreement with earlier numerical 
works$^{10}$ by  localization-length scalings. Therefore, various known 
localization-delocalization transitions in 2D non-interacting systems now can
be all described in terms of a topological characterization. Whether 
there is a far-reaching physical implication behind it needs to be further
studied.  
 
{\bf Acknowledgments} -The authors would like to thank T. K. Lee  and C. S. 
Ting for helpful discussions. The 
present work is supported  by the Texas Center for Superconductivity at the 
University of Houston and the Texas Advanced Research Program under Grant no 
3652182.

\newpage  
 
\noindent $^1$ E. Abrahams, P. W. Anderson, D. C. Licciardello, and V.  
Ramakrishnan, Phys. Rev. Lett. {\bf 42}, 673 (1979).\\ 
$^2$ For a review, see, P. A. Lee and T. V. Ramakrishnan, Rev. Mod. Phys. 
${\bf 57}$, 287 (1985).\\ 
$^3$ R. B. Laughlin, Phys Rev. B. ${\bf 23}$, 5632 (1981); B. I. Halperin, 
Phys. Rev. B. ${\bf 25}$, 2185 (1982).\\ 
$^4$ H. Levine, S. B. Libby and A. M. M. Pruisken, Phys. Rev. Lett. ${\bf 51}$,
1915 (1983); A. M. M. Pruisken, in {\bf The   Quantum  Hall  Effect}, edited 
by R. E. Prange and S. M. Girvin ( Springer-Verlag, Berlin, 1990).\\ 
$^{5}$ S. C. Zhang and D. Arovas, Phys. Rev. Lett. {\bf 72}, 
1886, (1994); V. Kalmeyer and S. C. Zhang, Phys. Rev. B{\bf 46}, 9889 (1992); 
Y. Avishai and Y. Hatsugai and M. Kohmoto, Phys. Rev. B. ${\bf 47}$, 9561 
(1993); V. Kalmeyer, D. Wei, D. P. Arovas, and  S. C. Zhang,
Phys. Rev. B. ${\bf 48}$, 11095 (1993); D. Z. Liu, X. C. Xie, S. Das Sarma and 
S. C. Zhang, preprint (1994).\\ 
$^{6}$ A. G. Aronov, A. D. Mirlin and P. Wolfle,  
 Phys. Rev. B ${\bf 49}$, 16609 (1994); 
 T. Sugiyama and N. Nagaosa, Phys. Rev. Lett. ${\bf 70}$, 1980 (1993); 
 D. Lee and J. Chalker,  Phys. Rev. Lett. ${\bf 72}$, 1510 (1994).\\ 
$^7$ D. N. Sheng and Z. Y. Weng, Phys. Rev. Lett. ${\bf 75}$, 2388 (1995).\\  
$^{8}$ F. Wegner, Nucl. Phys. B.  ${\bf 316}$, 663 (1989).\\ 
$^{9}$ S. Hikami, A. I. Larkin and Y. Nagaoka, Prog. Theor. Phys.  
 ${\bf 63}$, 707 (1980);  S. Hikami, J. Physique Lett. 
 ${\bf 46}$, L719 (1985).\\ 
$^{10}$  T. Ando, Phys. Rev. B {\bf 40}, 5325 (1989); 
A. MacKinnon, The scaling theory of localisation, in: Localization, 
Interaction and Transport Phenomena, B. Kramer, G. Bergmann and Y. 
Bruynseraede, eds., Berlin, 1985) 90; 
U. Fastenrath et al., Phisica A   ${\bf 172}$, 302 (1991);  
U. Fastenrath, Phisica A   ${\bf 189}$, 27 (1992);  
S. N. Evangelou and T. Ziman, J. Phys. C. ${\bf 20}$, L235 (1987); 
S. N. Evangelou, Phys. Rev. Lett. ${\bf 75}$, 2550 (1995).\\
$^{11}$ F. J. Dyson, J. Math. Phys. ${\bf 3}$, 140 (1962); ${\bf 3}$, 
 157 (1962);    ${\bf 3}$, 166 (1962);   ${\bf 3}$, 1191 (1962); 
 ${\bf 3}$, 1199 (1962).\\ 
$^{12}$ D. J. Thouless, M. Kohmoto, M. P. Nightingale, and M. den Nijs, Phys. 
Rev. Lett. ${\bf 49}$, 405 (1982);  D. J. Thouless, J. Phys. C, {\bf 17}, L325 
(1984); Q. Niu, D. J. Thouless, and Y. S. Wu, Phys. 
Rev. B {\bf 31}, 3372 (1985).\\ 
$^{13}$ M. Kohmoto, Ann. Phys. (NY) ${\bf 160}$, 343 (1985).\\ 
$^{14}$ D. P. Arovas et. al., Phys. Rev. Lett. ${\bf 60}$, 619 (1988).\\ 
$^{15}$ Y. Huo and R. N. Bhatt, Phys. Rev. Lett. ${\bf 68}$, 1375 (1992).\\ 
$^{16}$ M. V. Berry, Proc. R. Soc. Lond. A. ${\bf 392}$, 45 (1984); 
 B. Simon, Phys. Rev. Lett. ${\bf 51}$, 2167 (1983).\\  
$^{17}$ A.H. MacDonald, Phys. Rev. B{\bf 29}, 3057 (1984); M. Nielsen and P.
Hedegard, Phys. Rev. B. ${\bf 51}$, 7679 (1995).\\ 
$^{18}$ K. Chaltikian, L. Pryadko, and S. C. Zhang, preprint, 
cond-mat/9411029.\\  
$^{19}$ D. N. Sheng and Z. Y. Weng, in progress.\\
$^{20}$ E. Medina and M. Kardar,  Phys. Rev. Lett. ${\bf 66}$, 3187 (1991).\\
$^{21}$ L. Schweitzer,  preprint, cond-mat (1995).\\ 
$^{22}$ J. T. Chalker and G. T. Daniell, Phys. Rev. Lett.    
${\bf 61}$, 593 (1988); B. Huckenstein and  L. Schweitzer, 
Phys. Rev. Lett.   ${\bf 72}$, 713 (1994).\\
$^{23}$ F. B. Mancoff et al., preprint (1994). \\ 
$^{24}$ Y. Shapir and Z. Ovadyahu, 
 Phys. Rev. B. ${\bf 40}$, 12441 (1989).\\ 
$^{25}$ S. Y. Hsu and J. M. Valles, Jr., Phys. Rev. Lett. ${\bf 74}$, 2331 
 (1995).\\ 
$^{26}$ G. Bergmann,  Phys. Rep.   ${\bf 1}$, 107 (1984).\\ 
  
\newpage 
\begin{center} 
{\bf FIGURE CAPTION} 
\end{center} 
 
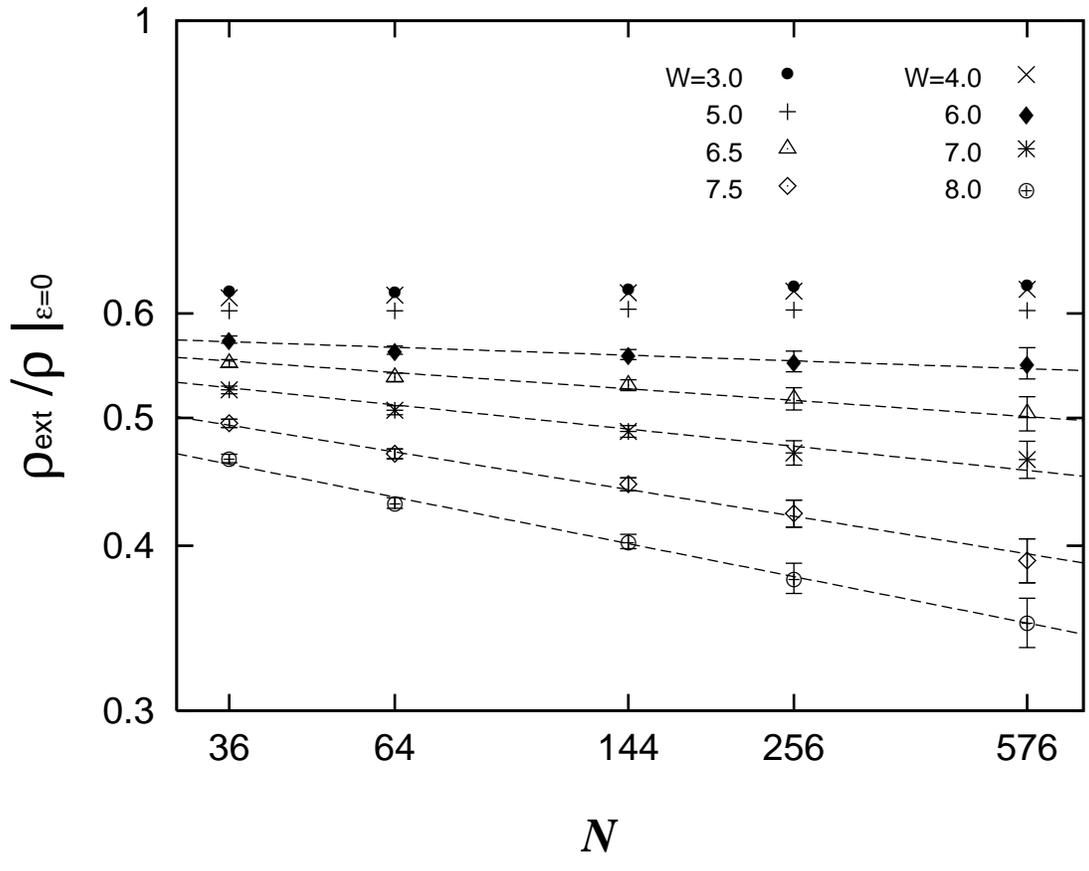
Fig. 1.  The ratio of the density of extended states over  the total  
density  of states   is  plotted  
as a function of sample size ${N}$ at energy  $\varepsilon=0$.
$W$ is the strength of random scattering. Error bars of the data for $W\leq 5.0
$ are in the same order of magnitude as those for $W> 5.0$ but are not marked 
in the figure for clarity.  \\

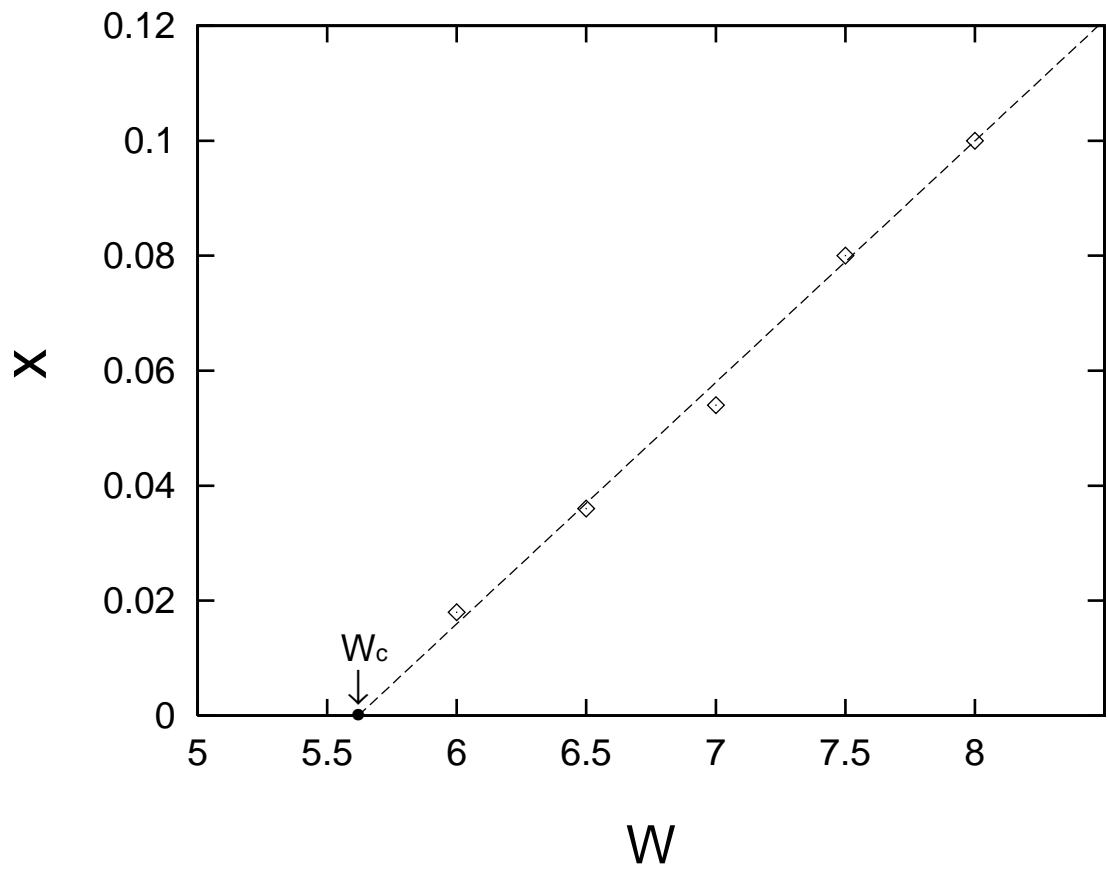
Fig. 2. The exponent $x$ vs. the strength of random scattering $W$. The data  
is extrapolated to $x=0$, giving a critical 
strength $W_c=5.62\pm 0.08$. 
\end{document}